\documentclass[12pt,a4paper]{JHEP3}

\usepackage{amscd,amsmath,amssymb,amsfonts,xspace,mathtools}
\usepackage{xcolor}
\usepackage{tikz-cd}
\usepackage{url}
\let\normalcolor\relax      
\usepackage{latexsym}  
\usepackage{wasysym}  
 \usepackage{graphicx}

\usepackage{graphicx, subfig, float}

\usepackage{epsfig,amsfonts,amssymb}
\usepackage{framed}

\newcommand{\refb}[1]{(\ref{#1})}
\newcommand{\Tr}{\mathop{\mathrm{Tr}}}

\newcommand{\nn}{\ensuremath{\nonumber{}}}

\title{Non-trivial saddles in microscopic description  of black holes}

\author{Pranav Kumar$^{1}$, Swapnamay Mondal$^{2,3}$
\\
{\it $~^1$ Capgemini Engineering, Kadubisanahalli, Bengaluru, 560103, India} \\
{\it $~^2$ Dublin Institute for Advanced Studies, 10 Burlington Road, Dublin, Ireland}\\
{\it $~^3$ School of Mathematics, Trinity College, Dublin 2, Ireland}
\vspace*{2mm}\\
{\tt pranavkr29@gmail.com, swapno@maths.tcd.ie }
\vspace*{-3mm}
}

\abstract{
Non-trivial gravitational saddles have played a key  role in the island proposal for the black hole information paradox. It is worth asking if non-trivial saddles exist in microscopic descriptions of black holes. We show this to be the case for 1/8 BPS black holes in  $\mathcal{N}=8$ String  Theory in a duality frame, where all charges are Ramond Ramond. 
The saddles are in the Coulomb branch, where they describe marginally stable bound states of the constituent branes,  and correspond to vacua of the BFSS model. The non-perturbative suppression scale is determined by the binding energy. 
\\

}

\begin{document}
 
\maketitle	

\section{Introduction}
Black hole information  paradox is about the apparent violation of unitarity in black hole spacetimes \cite{Hawking:1975vcx, Hawking:1976ra}. 
In terms  of entanglement entropy of Hawking radiation, the paradox is that unitarity would demand the final state of the radiation to have vanishing entropy, but Hawking's computation suggests otherwise. The puzzle is - how can the entropy come down?

Since the paradox is tied to quantum gravity, AdS/CFT  \cite{Maldacena:1997re, Witten:1998qj} offers a useful setting to study the paradox.  
 In holographic spacetimes, entanglement entropy can be studied using the Ryu-Takayanagi formula \cite{Ryu:2006bv}, and its generalizations \cite{Hubeny:2007xt,Faulkner:2013ana}. In particular, Von Neumann entropy of an evaporating black hole can be computed using the minimal quantum extremal surface \cite{Engelhardt:2014gca}.

It was shown in  \cite{Penington:2019npb, Almheiri:2019psf} that for an old black hole, i.e. one past the Page time, the quantum extremal surface lies slightly behind the event horizon. This reproduces the correct qualitative behaviour of the Page curve \cite{Page:1993wv, Page:2013dx}. Non-perturbative errors in the entanglement wedge reconstruction \cite{Czech:2012bh, Headrick:2014cta, Wall:2012uf} play an important role in this argument.

However, the information paradox really  concerns the Page curve of the radiation.
In \cite{Almheiri:2019hni},  the entropy of Hawking radiation was computed using the quantum extremal surface prescription in a holographic set up. Black hole interior was found to be part of the entanglement wedge of the radiation.  This led to the so-called island proposal, which suggests that for Von Neumann entropy of quantum systems coupled to gravity, one has to include the ``quantum extremal islands'', which are regions in gravity theory entangled with external quantum matter.
From this perspective, information paradox arises because we have been using the wrong entropy formula.
The key insight towards the proposed resolution is that non-perturbative effects play an important role.

Island proposal for various black holes have been investigated in 
\cite{Hashimoto:2020cas, Anegawa:2020ezn, Ling:2020laa, Karananas:2020fwx, Goto:2020wnk, Wang:2021woy, Lu:2021gmv, Yu:2021rfg, Ahn:2021chg, Anand:2023ozw, Gan:2022jay, Goswami:2022ylc, Emparan:2023dxm, Emparan:2021hyr, Jeong:2023hrb}. In cosmological context, island proposal has been studied in \cite{Espindola:2022fqb, Bousso:2022gth, VanRaamsdonk:2020tlr, Hartman:2020khs, Azarnia:2021uch}. 
For criticisms of island proposal, see \cite{Geng:2020qvw, Geng:2021hlu, Li:2021lfo, Karlsson:2020uga, Karlsson:2021vlh, Singh:2022ljh, Raju:2021lwh}. Other relevant works include \cite{Almheiri:2018xdw, Hayden:2018khn, Kundu:2021nwp, Caceres:2020jcn}.

Von Neumann entropy, the main object of the island proposal, can be extracted from the $n \to 1$ limit of R\'{e}nyi entropy $\Tr{} \rho^n$, which can be thought of as an observable on $n$ copies of the relevant system \cite{Callan:1994py}. Building on earlier works  \cite{Lewkowycz:2013nqa, Dong:2016hjy, Dong:2017xht}, it was shown in \cite{Almheiri_2020, Penington:2019kki} that the computation of R\'{e}nyi entropy receives corrections from non-trivial gravitational saddles, dubbed replica wormholes. It was further shown that in $n \to 1$ limit, these corrections lead to the island proposal. The moral of the story is that non-trivial saddles play a crucial role in black hole physics.  

It was noted in \cite{Almheiri_2020} itself that the island proposal is not a full resolution of the information paradox, as the Hilbert space on which the supposed unitary time evolution acts, is missing. This is somewhat complementary to microscopic D-brane descriptions of black holes, where microstates are apparent \cite{Strominger:1996sh, Maldacena:1997de}. This prompts us to ask - {\it is there a manifestation of the island proposal in microscopic descriptions of black holes?}

One would think no, firstly because microscopic descriptions are explicitly unitary, hence there is no paradox to be resolved and secondly because they lack gravity\footnote{Some progress towards incorporating gravity in conformal field theory descriptions has recently been achieved in \cite{PhysRevD.107.L121903}.}. Nevertheless one might ask - is there a qualitative similarity between the island picture and microscopic descriptions of black holes? This would amount to asking- are there non-trivial saddles in the microscopic descriptions of black holes? 

In this work, we show this to be the case for certain microscopic descriptions of 1/8 BPS black holes in $\mathcal{N}=8$ String theory. The low energy dynamics of the D-brane system is that of an effective particle.
The microscopic Lagrangian is known from \cite{Chowdhury:2014yca, Chowdhury:2015gbk}. In this system, we find the potential has a non-compact critical manifold. The critical points correspond to the vacua of four sets of Banks-Fischler-Shenker-Susskind (BFSS) models \cite{Banks:1996vh} and can be interpreted as marginal bound states of the constituent D-branes. 

The paper is organized as follows. In Section \ref{d2-d2-d2-d6}, we discuss the relevant D-brane system. In Section \ref{saddles}, we discuss the non-trivial saddles of this system. In Section \ref{sdisc}, we comment on possible implications of our findings.
\section{Microscopic description of the $D2 \textendash D2 \textendash D2 \textendash D6$ black hole} \label{d2-d2-d2-d6}
We consider 4-charge 1/8 BPS black holes in $\mathcal{N}=8$ String theory in four dimensions. 
The entropies of these black holes were studied using quantum entropy function in \cite{Sen:2009gy}. A microscopic computation of entropy  was performed in \cite{Shih:2005qf}.
The relevant index is the 14-th helicity  supertrace \cite{Bachas:1996bp, Gregori:1997hi}
\begin{align}
B_{14} &= - \frac{1}{14 !}  \Tr{} (2J_3)^{14} (-1)^{2J_3} \, ,
\end{align}
which is same as the Witten index after removing the Goldstino multiplets. Restricting to configurations with charge vector $(1,1,1,N)$, the generating function for $B_{14}(N)$ is given by\footnote{First few terms in the generating function are $2+12 q+56 q^2+208 q^3+684 q^4+2032 q^5+5616 q^6+14592 q^7+36088 q^8+85500 q^9+195312 q^{10}+ \dots \, .$} 
\begin{align}
\sum_{N=0}^\infty B_{14}(N) q^N &= \frac{\sum_{n \in \mathbb{Z}} q^{n(n+1)}}{ \prod_{n=1}^\infty (1-q^n)^6} \, .
\end{align}
In order to compute this index, \cite{Shih:2005qf} made use of the 4d-5d connection \cite{Gaiotto:2005gf} and known microscopic entropies of the corresponding five dimensional black holes \cite{Maldacena:1999bp}. Ref. \cite{Maldacena:1999bp} considered a D1-D5-momentum system, whose low energy physics is described by an effective string. The world sheet theory of this effective string is captured by a two dimensional conformal field theory (CFT). Note, this CFT captures a family of four dimensional black holes at one go. In order to study the dynamics of a black hole with a fixed charge, one can restrict to  that specific charge sector in the CFT. A more direct route however would be to construct microscopic descriptions for individual charge sectors. 

This was achieved in \cite{Chowdhury:2014yca, Chowdhury:2015gbk} for 4-charge 1/8 BPS black holes, in a reference frame where all charges are Ramond-Ramond charges. The resulting microscopic system comprises of four stacks of D-branes: three D2 branes wrapping three disjoint two-cycles inside the compact six-torus, and a stack of $N$ D6 branes wrapping the entire six-torus. The brane configuration is summarized in  Table \ref{Dconfig}. 
\begin{table}[htb]
\begin{center}
\begin{tabular}{|c||c|c|c|c|c|c|c|c|c|c}
\hline
& 1 & 2 & 3 & 4 & 5 & 6 & 7 & 8 & 9 \\
\hline
D2 &  &  &  & $\checked$ & $\checked$ &  &  &  &  \\
\hline
D2 &  &  &  &  &  & $\checked$ & $\checked$ &  &  \\
\hline
D2 &  &  &  &  &  &  &  & $\checked$ & $\checked$ \\
\hline
D6 &  &  &  & $\checked$ & $\checked$ & $\checked$ & $\checked$ & $\checked$ & $\checked$ \\
\hline
\end{tabular}
\end{center}
\caption{The D-brane configuration}\label{Dconfig}
\end{table}
First three stacks each break half of the available supersymmetry, altogether preserving 4 supercharges, which amounts to $\mathcal{N}=1$ supersymmetry in four dimensions. It is possible to choose the orientation of the fourth brane in a manner so as to preserve these 4 supercharges. The four stacks intersect at a point, thus the low energy dynamics is that of an effective (super)particle. The supersymmetric quantum mechanics of this effective particle was explicitly constructed in \cite{Chowdhury:2014yca, Chowdhury:2015gbk}, and was shown to reproduce the correct degeneracies for small charges\footnote{ For $N>3$, the tools used in \cite{Chowdhury:2014yca, Chowdhury:2015gbk} proved insufficient. See \cite{Maji:2023ims} for some recent progress.}. 

The detailed Lagrangian of the system is discussed in \cite{Chowdhury:2014yca, Chowdhury:2015gbk}, and more recently in \cite{Kumar:2023hlu}. Here we gather only  what is needed for our purpose. 

Each stack preserves 16 supercharges and is described by a $\mathcal{N}=4$ super Yang Mills theory \cite{Witten:1995im}, reduced to $0+1$ dimension. This can also be thought of as BFSS  model \cite{Banks:1996vh}. Each pair of branes preserves 8 supercharges, hence the interaction of each pair can be described by dimensional reduction of a $\mathcal{N}=2$ Lagrangian in four dimensions. Different pairs preserve different $\mathcal{N}=2$ subalgebras, altogether preserving only a  $\mathcal{N}=1$ sub-algebra,  i.e. 4 supercharges.
As explained above, the entire system preserves 4 supercharges, allowing for a  further addition of a superpotential as well as Fayet-Iliopoulos (FI) parameters \cite{Fayet:1974jb}. 

The bosonic fields of $\mathcal{N}=4$ super Yang Mills for the $k^{th}$ stack comprise the real fields $X^{k}_1, \, X^{k}_2, \, X^{k}_3$ (describing three  non-compact spatial coordinates) and the complex fields  $\Phi^{k}_1, \, \Phi^{k}_2, \, \Phi^{k}_3$ (describing compact spatial coordinates along the three two-tori).  
This covers 9 spatial coordinates. Fields with superscript $k=1,2,3$ are numbers and those with superscript $4$ are $N \times N$ matrices. 
Apart from these, there are two complex  scalars $Z^{kl}, Z^{lk}$, coming from a  $\mathcal{N}=2$ hypermultiplet for every pair $(kl)$ of stacks , $k,l=1,2,3,4$. All these bosons are accompanied by their fermionic partners. For further details, we refer the reader to \cite{Chowdhury:2014yca, Chowdhury:2015gbk}. 
 
Now we come to the potential. This has recently been discussed in \cite{Kumar:2023hlu}. It is a sum of three non-negative terms
\begin{align}
V &=  V_{gauge} +V_D + V_F  \, , \label{Vtotal}
\end{align}
with
\begin{align}
\nn
V_{gauge} &=  \sum_{k,l=1; k\neq l}^4 \sum_{a=1}^3 \Tr{} \left[ \left( H^{kl}_a \right)^\dagger \left( H^{kl}_a \right) \right] +   \sum_{k=1}^4 \sum_{a=1}^3 \sum_{A=1}^3 \Tr{} \left[ \left( Y^{k}_{aA} \right)^\dagger Y^{k}_{aA} \right] \\
\nonumber
&+ \frac{1}{4} \sum_{k=1}^4 \sum_{a,b=1}^3 \Tr{} \left[ \left( X^{k}_{ab} \right)^\dagger X^{k}_{ab}  \right] \, , \\
\nn
V_D &= \frac{1}{2} \sum_{k=1}^4 \Tr{} \left[ \left( \sum_{l \neq k} Z^{kl} Z^{kl \dagger} -  \sum_{l \neq k} Z^{lk \dagger} Z^{lk} + \sum_{A=1}^3 \left[ \Phi^{k}_A, \Phi^{k\dagger}_A \right] - c^{(k)} \mathbb{I}_{N_k} \right)^2 \right] \, , \\
\nn
V_F &= 2 \Big( \sum_{i \neq j, i,j =1}^3 |F^{ij}|^2 + \sum_{i=1}^3 |F^{i4}|^2 +  \sum_{i=1}^3 \Tr{} (F^{4i})^\dagger F^{4i} \\
&+ \sum_{i \neq j, i,j =1}^3 |G^{ij}|^2 + \sum_{i=1}^3 \left( (G^{4i})^\dagger G^{4i} + G^{i4} (G^{i4})^\dagger  \right) \label{V_F} \Big) \, . 
\end{align}
Here $c^{(k)}$-s are FI parameters, satisfying $c^{(1)} + c^{(2)} + c^{(3)} + N c^{(4)} = 0$.
Coming  to $V_{gauge}$,
\begin{align}
X^{k}_{ab} &= [X^{k}_a, X^{k}_b]\, , 
Y^{k}_{aA} = [X^{k}_a,\Phi^{k}_A] \, , H^{kl}_a = X_a^{k} Z^{kl} - Z^{kl} X_a^{l} \, .
\end{align}
In $V_F$, 
\begin{align}
\nn
F^{12} &=  Z^{12} Z^{21} + c^{12}   ~,~
F^{21} =  Z^{21} Z^{12} + c^{12} , \\
\nn
F^{23} &= Z^{23} Z^{32} + c^{23}  ~,~ 
F^{32} =  Z^{32} Z^{23} + c^{23}  , \\
\nn
F^{13} &=  Z^{13} Z^{31} + c^{13}   ~,~
F^{31} =  Z^{31} Z^{13} + c^{13}  , \\
\nn
F^{14} &= Z^{14}. Z^{41} + c^{14} N  ~,~
F^{41} =  Z^{41} Z^{14} + c^{14} \mathbb{I}_{N} + [\Phi_2^4, \Phi_3^4]  , \\
\nn
F^{24} &=  Z^{24}. Z^{42} + c^{24} N ~,~ 
F^{42} =  Z^{42} Z^{24} + c^{24} \mathbb{I}_{N} + [\Phi_3^4, \Phi_1^4]  , \\
\nn
F^{34} &=   Z^{34}. Z^{43} + c^{34} N  ~,~
F^{43} =  Z^{43} Z^{34} + c^{34} \mathbb{I}_{N} - [\Phi_2^4,\Phi_1^4] , \\
\nn
G^{21} &=  \Phi^{(12)}  Z^{21} + Z^{23} Z^{31} + Z^{24} . Z^{41}  ~,~ G^{12} =  \Phi^{(12)}  Z^{12} + Z^{13} Z^{32} + Z^{14} . Z^{42}  \, , \\
\nn
G^{31} &=  \Phi^{(31)} Z^{31}  + Z^{32} Z^{21} - Z^{34} . Z^{41}   ~,~ G^{13} =  \Phi^{(31)}  Z^{13}  + Z^{12} Z^{23} + Z^{14} . Z^{43} \, ,  \\
\nn
G^{32} &=  \Phi^{(23)} Z^{32}  + Z^{31} Z^{12} + Z^{34} . Z^{42}  ~,~ G^{23} = \Phi^{(23)} Z^{23} + Z^{21} Z^{13} + Z^{24} . Z^{43}  \, ,  \\
\nn
G^{41} &=   - \Phi_1^4 Z^{41} + Z^{42} Z^{21} + Z^{43} Z^{31}  ~,~ G^{14} =  - Z^{14} \Phi_1^4 + Z^{12} Z^{24} - Z^{13} Z^{34}  \, ,  \\
\nn
G^{42} &=  - \Phi_2^4 Z^{42} + Z^{43} Z^{32} + Z^{41} Z^{12}  ~,~ G^{24} = - Z^{24} \Phi_2^4 + Z^{21} Z^{14} + Z^{23} Z^{34}  \, ,  \\
G^{43} &=  - \Phi_3^4 Z^{43} - Z^{41} Z^{13} + Z^{42} Z^{23}   ~,~ G^{34} =  - Z^{34} \Phi_3^4 + Z^{31} Z^{14} + Z^{32} Z^{24}  \, . \label{Fterm}
\end{align}
Here 
\begin{align}
\Phi^{(12)} &:= \Phi^1_3 - \Phi^2_3 \, , \, \Phi^{(23)}  = \Phi^2_1 - \Phi^3_1 \, , \, \Phi^{(31)} :=  \Phi^3_2 - \Phi^1_2 \, , 
\end{align}
and $c^{12}, c^{13}, c^{23}, c^{14} , c^{24} , c^{34}$ are complex non-zero parameters. Note that $F^{41}, F^{42}, F^{43}$ are matrices, $G^{41}, G^{42}, G^{43}$ are column vectors, $G^{14}, G^{24}, G^{34}$ are row vectors and rest of the combinations are numbers. 
The parameters $c^{(k)}, c^{kl}$-s are related values of the metric and B-field, as described in  \cite{Chowdhury:2014yca, Chowdhury:2015gbk}. 

The potential \refb{Vtotal}  is invariant under the following gauge transformations
\begin{align}
\nonumber
&Z^{12} \rightarrow \lambda_{12} Z^{12} \, , \, Z^{21} \rightarrow \lambda_{12}^{-1} Z^{21} \, , \, 
Z^{23} \rightarrow \lambda_{23} Z^{23} \, , \, Z^{32} \rightarrow \lambda_{23}^{-1} Z^{32} \, , \\
\nonumber
&Z^{13} \rightarrow \lambda_{12} \lambda_{23} Z^{13} \, , \, Z^{31} \rightarrow \lambda_{12}^{-1} \lambda_{23}^{-1} Z^{31} \, , \\
\nonumber
&Z^{14} \rightarrow \lambda_{12} \lambda_{23} Z^{14} M^{-1} \, ,  Z^{24} \rightarrow \lambda_{23} Z^{24} M^{-1} \, ,  Z^{34} \rightarrow  Z^{34} M^{-1} \, , \\
\nonumber
&Z^{41} \rightarrow \lambda_{12}^{-1} \lambda_{23}^{-1} M Z^{41} \, , \, Z^{42} \rightarrow \lambda_{23}^{-1} M Z^{42} \, , \, Z^{43} \rightarrow M Z^{43} \, , \\
&\Phi_i^{4} \rightarrow M \Phi^{4}_i M^{-1} \, , \label{complexgauge}
\end{align}
where $\lambda_{12} \in U(1), \lambda_{23} \in U(1),  M \in U(N)$. The zeroes of  $V_F$ are  invariant under a larger $\mathbb{C}^\ast \times \mathbb{C}^\ast \times GL(N, \mathbb{C})$ symmetry, which  was exploited in \cite{Chowdhury:2014yca, Chowdhury:2015gbk,Kumar:2023hlu}. The D-term potential can be traded for this enhancement of gauge symmetry, as far as supersymmetric minima are concerned. 
\section{Saddles of the $D2 \textendash D2 \textendash D2 \textendash D6$ black hole} \label{saddles}
Different D-brane stacks interact through strings stretched between those stacks. One might therefore expect the ``unstrung stacks" to be as free as say a stack of D3 branes or  a bunch  of D0  branes. In the following, we show that  indeed such unstrung configurations arise as a continuum of saddles for the potential \refb{Vtotal}.

Firstly, we note that in the absence of the $Z$ fields, different D-brane stacks do not interact and the Lagrangian is a sum of four  Lagrangians, each describing one stack. These individual Lagrangians are dimensional reductions of $\mathcal{N}=4$ Super Yang Mills \cite{Witten:1995im} or the BFSS model \cite{Banks:1996vh} except that
\begin{enumerate}
\item
there are non-zero FI parameters $c^{(k)}$-s (which enter $V_D$) and
\item
there is a superpotential that is linear in $\Phi$, involving the parameters $c^{ij}$-s (which enter $V_F$ as the constant  terms in $F^{ij}$-s).
\end{enumerate}
In $V_D$, the linear terms in the FI parameters vanish for vanishing $Z$-fields, whereas the quadratic term is a constant. So the only change in $V_D$ in this case is the addition of a constant term. 
In $V_F$, the linear terms in $c^{ij}$-s vanish for vanishing  $Z$-s, and the only change is a constant, quadratic in the $c^{ij}$-s. 

Thus we see that for vanishing $Z$ fields, up to some additive positive constants, the potential for each stack is that of a dimensionally reduced $\mathcal{N}=4$ Super Yang Mills, or equivalently the BFSS model \cite{Banks:1996vh}. It follows  that, in the subspace with vanishing $Z$ fields, the potential minima of BFSS model, which are known to be configurations with commuting $\Phi$ and $X$ fields, are the local minima of the system at hand for $X$ and $\Phi$ directions. 
Note that in the present case, these are local extrema and not global minima.

It remains to be checked whether or not  these configurations are extrema with respect to Z-s as well. We make the simple observation that there are no terms in the potential which are linear in  the $Z$-s. 
It  follows that derivative of the potential with respect to any $Z$ contains only positive powers of $Z$-s, and thereby vanishes for vanishing $Z$-s. Thus we reach the conclusion that vanishing $Z$ fields and commuting $\Phi$ and $X$ fields define a critical manifold $\mathcal{M}_{crit}$ for the potential. 
The value of the potential on this critical manifold is given by 
 \begin{align}
 V_{crit} &= \frac{1}{2} \left( (c^{(1)})^2 + (c^{(2)})^2 + (c^{(3)})^2 + N (c^{(4)})^2 \right) + 4 \sum_{k<l; k,l =1}^3 |c^{kl}|^2 + 4 N^2 \sum_{k=1}^3 |c^{4k}|^2 \, .
 \end{align}
 Note, $\mathcal{M}_{crit}$ is non-compact, thus it is possible to construct non-normalizable states supported on $\mathcal{M}_{crit}$. In this sense, these states are like scattering states. 
Physically, it  seems fit to interpret the configurations on $\mathcal{M}_{crit}$ as marginally stable bound states of the constituent branes. Reaching the true bound  states, associated with the global minima (with vanishing $V$), then saves $V_{crit}$ amount of energy. This suggests $V_{crit}$ should  be thought of as binding energy of the system.  
  
On the critical manifold $\mathcal{M}_{crit}$, the gauge symmetry is broken to a product of $U(1)$-s. Thus $\mathcal{M}_{crit}$ belongs to the Coulomb branch of this brane system. Another feature of the Coulomb branch in quiver  quantum mechanics  is the absence of bifundamentals. This feature is also exhibited in the present case.

What remains to be explored is the nature of the extrema. We do not study this question in this paper, with the exception of the $N=1$ case, which is discussed in Appendix \ref{appHess}. We find the extrema are minima when the constituent branes are not too close. When the constituent branes come closer than a certain distance, unstable directions open up. This distance is set by $c^{(i)}$-s and $c^{ij}$-s, and vanishes for vanishing $c^{(i)}$-s and $c^{ij}$-s, turning points on the critical manifold into true minima. In the same case, $V_{crit}$ vanishes, hence turning $\mathcal{M}_{crit}$ into the vacuum manifold. This gives insight into why the vacuum manifold is non-compact for vanishing $c^{(i)}, c^{ij}$-s.

In $N=1$ case, we also numerically  confirmed the existence of these minima, using stochastic gradient descent algorithm within Tensorflow library \cite{tensorflow2015-whitepaper}, treating the potential as the loss function \cite{Kumar:2023hlu}. Also, our numerical search did not reveal any other minima.

Now we come to the corresponding saddles in the path integral. Points on the critical manifold lead to saddles with infinite action (due to the time integral). In comparison, the gravitational wormholes usually have finite action\footnote{We thank Ashoke Sen for stressing this point.}.  Nevertheless in the present case, saddles will modify the various correlation functions, as computed from the leading saddles corresponding to the global minima.
In fact, given the non-compactness of the critical manifold, such modifications may be quite significant.
Finiteness of the saddle actions aside, there is a commonality with the island picture in that the non-trivial saddles give significant contributions in both cases. 
\section{Discussion} \label{sdisc}
In this paper, we have illustrated the  existence of a non-compact critical manifold in microscopic description of 1/8 BPS black holes in $\mathcal{N}=8$ String Theory. Although the contribution of corresponding saddles are exponentially suppressed, cumulatively they are likely to give significant contributions to various correlation functions due to the non-compactness of the critical manifold. It is beyond the scope of  this paper to estimate such contributions. The suppression scale is determined by the binding energy $V_{crit}$. Configurations on the critical manifold can be thought of as marginally stable bound states of the constituent D-branes.

The analysis of this paper is based on the Lagrangian used in \cite{Chowdhury:2014yca, Chowdhury:2015gbk}. 
However, higher point amplitudes in the string worldsheet will lead to corrections to this Lagrangian.
 It is clear that higher order terms in $Z$-s will not change  these extrema. As  for higher order terms in $\Phi$-s and $X$-s, we do not have a rigorous argument for similar conclusion. But we note the commuting  $\Phi$-s and $X$-s lead to a rather intuitive interpretation of the minima. It would be strange if this interpretation is lost due to higher order terms.   

The gravity manifestation of these saddles is not clear.
Main hurdle in making progress in this direction is our limited understanding of incorporation of gravity in microscopic descriptions. The arguments of \cite{PhysRevD.107.L121903}, which suggests $T\overline{T}$ deformation captures the effect of gravity in microscopic description, are limited to conformal field theory descriptions. 
Nevertheless, taking a cue from the simple physical interpretation of the critical manifold, we can ask if there are marginally stable bound states of the infalling matter that formed the black hole.

The existence of the critical manifold is quite interesting for the brane system itself. For example, it is curious to note that the zero angular momentum conjecture \cite{Chowdhury:2015gbk}, which in this setup means that the supersymmetric minima form a zero dimensional space, holds whenever there are non-trivial saddles. We are not sure if this is a coincidence or if there is more to it.

Also, non-compactness of the vacuum manifold at special values of the moduli (vanishing $c^{(k)}, c^{kl}$-s) is seen to be a result of the critical manifold merging with the discreet vacua. It is worth investigating whether similar picture arises in other supersymmetric problems with non-compact vacuum manifolds. One such system would be the Coulomb branch of a scaling quiver. 

We conclude by noting that it would be interesting to compute the contribution of these saddles to various correlation functions.  

\paragraph{Acknowledgement:} We would like to thank Ashoke Sen for his comments.

\appendix
\section{Nature of $\mathcal{M}_{crit}$ in $N=1$ case}\label{appHess}
For the Abelian case, all variables are numbers, leading to considerable simplification. In this case we have
\begin{align}
\nn
V_D &= \frac{1}{2} \sum_{k=1}^4  \left( \sum_{l \neq k} | Z^{kl} |^2  -  \sum_{l \neq k} | Z^{lk} |^2  - c^{(k)} \right)^2 \, , \\
\nonumber
V_{gauge} &=  \sum_{k,l=1; k\neq l}^4 \sum_{a=1}^3  \left( X_a^{k} -  X_a^{l}  \right)^2  \left|  Z^{kl}  \right|^2   \, , \\
V_F &= 2 \Big( \sum_{i \neq j, i,j =1}^4 |F^{ij}|^2 + \sum_{i \neq j, i,j =1}^4 |G^{ij}|^2  \Big) \, , 
\end{align}
where $F^{ij}, G^{ij}$-s are as given in \refb{Fterm}, with the commutators put to zero. Various second derivatives are 
\begin{align}
\nn
\frac{\partial^2 V_D}{\partial Z^{kl} \partial (Z^{kl})^*} \Big|_{Z=0} &=   c^{(l)} - c^{(k)} \, , \,
\frac{\partial^2 V_{gauge}}{\partial Z^{kl} \partial (Z^{kl})^*} \Big|_{Z=0} = (\vec{X}^{k} - \vec{X}^{l})^2 \, ,\\
\frac{\partial^2 V_F}{\partial Z^{kl} \partial Z^{lk} } 
\Big|_{Z=0} &= (c^{kl})^* \, , \,
\frac{\partial^2 V_F}{\partial (Z^{kl})^* \partial (Z^{lk})^* } 
\Big|_{Z=0} = c^{kl} \, , \,
\frac{\partial^2 V_F}{\partial Z^{kl} \partial (Z^{kl})^* } 
\Big|_{Z=0} = |\Phi^{(kl)} |^2 \, ,
\end{align}
where  $ \Phi^{(4k)} = - \Phi^{(k4)}  = \Phi^4_k $. Overall, we have
\begin{align}
\nn
\frac{\partial^2 V}{\partial Z^{kl} \partial (Z^{kl})^*} \Big|_{Z=0} &=  ( c^{(l)} - c^{(k)}) +   (\vec{X}^{k} - \vec{X}^{l})^2 + 2  |\Phi^{(kl)} |^2 =: ( c^{(l)} - c^{(k)}) + |r^{(kl)}|^2 \, ,
\\
\frac{\partial^2 V}{\partial Z^{kl} \partial Z^{lk} } \Big|_{Z=0} &= (c^{kl})^* \, , \,
\frac{\partial^2 V}{\partial (Z^{kl})^* \partial (Z^{lk})^* } \Big|_{Z=0} = c^{kl} \, ,
\end{align}
where $|r^{(kl)}|^2 :=  (\vec{X}^{k} - \vec{X}^{l})^2 + 2  |\Phi^{(kl)} |^2 $.

Let $Z^{kl} = a^{kl} + i b^{kl} $.
Thus
\begin{align}
\nn
\frac{\partial^2 V}{\partial (a^{kl})^2} &=  
( c^{(l)} - c^{(k)}) + |r^{(kl)}|^2 \, , \,
\frac{\partial^2 V}{\partial (b^{kl})^2} = 
 ( c^{(l)} - c^{(k)}) + |r^{(kl)}|^2 \,  , \\
\frac{\partial^2  V}{\partial a^{kl} \partial b^{kl}} &= 0 \, , \,
\frac{\partial^2 V}{\partial a^{kl} \partial a^{lk}} = 
2  \Re{} (c^{kl}) \, , \,
\frac{\partial^2 V}{\partial b^{kl} \partial b^{lk}} = 
 - 2  \Re{} (c^{kl}) \, , \,
\frac{\partial^2 V}{\partial a^{kl} \partial b^{lk}} =  
2  \Im{} (c^{kl}) \, .
\end{align}
$\Re{}$ and $\Im{}$ respectively stand for real and imaginary parts of a  complex number. We  note that the Hessian matrix is block diagonal with each $(kl)$ pair forming a block. In the basis $(a^{kl}, b^{kl} , a^{lk}, b^{lk} )$, the Hessian matrix is
\begin{align}
\frac{1}{2}
\begin{pmatrix}
 ( c^{(l)} - c^{(k)}) + |r^{(kl)}|^2 & 0 & \Re{} (c^{kl}) &  \Im{} (c^{kl}) \\
0 &  ( c^{(l)} - c^{(k)}) + |r^{(kl)}|^2 &  \Im{} (c^{kl})  & - \Re{} (c^{kl}) \\
\Re{} (c^{kl}) &  \Im{} (c^{kl})  &  ( c^{(k)} - c^{(l)}) + |r^{(kl)}|^2 & 0 \\
 \Im{} (c^{kl}) & -\Re{} (c^{kl}) & 0 &  ( c^{(k)} - c^{(l)}) + |r^{(kl)}|^2 \\
\end{pmatrix} \, .
\end{align}
The  eigenvalues are doubly degenerate and are given by
\begin{align}
\lambda_\pm &= \frac{1}{2} \left( |r^{(kl)}|^2 \pm \sqrt{ (c^{(k)} - c^{(l)})^2 + |c^{kl}|^2} \right) \, .
\end{align}
Thus, for  $|r^{(kl)}|^2 > \sqrt{ (c^{(k)} - c^{(l)})^2 + |c^{kl}|^2} $, all four eigenvalues are positive, whereas $|r^{(kl)}|^2 < \sqrt{ (c^{(k)} - c^{(l)})^2 + |c^{kl}|^2}$, there are two negative eigenvalues. This means the points on the critical manifold are mostly minima, until $r^{(kl)}$ drops below a certain value, when two unstable directions open up. This story could be repeated for all pairs $(kl)$.
Altogether, we get the picture that  when the constituent branes come too close, instabilities develop. 

\bibliography{island} 
\bibliographystyle{JHEP}   

\end{document}